\documentclass[final,3p,times,twocolumn,longtitle]{elsarticle}

\usepackage{lineno}
\usepackage{graphicx}
\usepackage{color}
\usepackage{dcolumn}
\usepackage{bm}
\usepackage{multirow,array,tabularx}

\usepackage{amssymb}
\usepackage{amsthm}
\usepackage{mathtools}

\journal{Nuclear Physics B}

\begin{document}

\begin{frontmatter}

\title{Evidence for the Onset of Color Transparency in $\rho^0$ Electroproduction off Nuclei}

\newcommand*{\ANL}{Argonne National Laboratory, Argonne, Illinois 60439}
\newcommand*{\ANLindex}{1}
\newcommand*{\UNH}{University of New Hampshire, Durham, New Hampshire 03824-3568}
\newcommand*{\UNHindex}{2}
\newcommand*{\UTFSM}{Universidad T\'{e}cnica Federico Santa Mar\'{i}a, Casilla 110-V Valpara\'{i}so, Chile}
\newcommand*{\UTFSMindex}{3}
\newcommand*{\JLAB}{Thomas Jefferson National Accelerator Facility, Newport News, Virginia 23606}
\newcommand*{\JLABindex}{4}
\newcommand*{\YEREVAN}{Yerevan Physics Institute, 375036 Yerevan, Armenia}
\newcommand*{\YEREVANindex}{5}
\newcommand*{\ASU}{Arizona State University, Tempe, Arizona 85287-1504}
\newcommand*{\ASUindex}{6}
\newcommand*{\UCLA}{University of California at Los Angeles, Los Angeles, California  90095-1547}
\newcommand*{\UCLAindex}{7}
\newcommand*{\CSUDH}{California State University, Dominguez Hills, Carson, CA 90747}
\newcommand*{\CSUDHindex}{8}
\newcommand*{\CANISIUS}{Canisius College, Buffalo, NY}
\newcommand*{\CANISIUSindex}{9}
\newcommand*{\CMU}{Carnegie Mellon University, Pittsburgh, Pennsylvania 15213}
\newcommand*{\CMUindex}{10}
\newcommand*{\CUA}{Catholic University of America, Washington, D.C. 20064}
\newcommand*{\CUAindex}{11}
\newcommand*{\SACLAY}{CEA, Centre de Saclay, Irfu/Service de Physique Nucl\'eaire, 91191 Gif-sur-Yvette, France}
\newcommand*{\SACLAYindex}{12}
\newcommand*{\CNU}{Christopher Newport University, Newport News, Virginia 23606}
\newcommand*{\CNUindex}{13}
\newcommand*{\UCONN}{University of Connecticut, Storrs, Connecticut 06269}
\newcommand*{\UCONNindex}{14}
\newcommand*{\EDINBURGH}{Edinburgh University, Edinburgh EH9 3JZ, United Kingdom}
\newcommand*{\EDINBURGHindex}{15}
\newcommand*{\FU}{Fairfield University, Fairfield CT 06824}
\newcommand*{\FUindex}{16}
\newcommand*{\FIU}{Florida International University, Miami, Florida 33199}
\newcommand*{\FIUindex}{17}
\newcommand*{\FSU}{Florida State University, Tallahassee, Florida 32306}
\newcommand*{\FSUindex}{18}
\newcommand*{\Genova}{Universit$\grave{a}$ di Genova, 16146 Genova, Italy}
\newcommand*{\Genovaindex}{19}
\newcommand*{\GWUI}{The George Washington University, Washington, DC 20052}
\newcommand*{\GWUIindex}{20}
\newcommand*{\ISU}{Idaho State University, Pocatello, Idaho 83209}
\newcommand*{\ISUindex}{21}
\newcommand*{\INFNFE}{INFN, Sezione di Ferrara, 44100 Ferrara, Italy}
\newcommand*{\INFNFEindex}{22}
\newcommand*{\INFNFR}{INFN, Laboratori Nazionali di Frascati, 00044 Frascati, Italy}
\newcommand*{\INFNFRindex}{23}
\newcommand*{\INFNGE}{INFN, Sezione di Genova, 16146 Genova, Italy}
\newcommand*{\INFNGEindex}{24}
\newcommand*{\INFNRO}{INFN, Sezione di Roma Tor Vergata, 00133 Rome, Italy}
\newcommand*{\INFNROindex}{25}
\newcommand*{\ORSAY}{Institut de Physique Nucl\'eaire ORSAY, Orsay, France}
\newcommand*{\ORSAYindex}{26}
\newcommand*{\ITEP}{Institute of Theoretical and Experimental Physics, Moscow, 117259, Russia}
\newcommand*{\ITEPindex}{27}
\newcommand*{\JMU}{James Madison University, Harrisonburg, Virginia 22807}
\newcommand*{\JMUindex}{28}
\newcommand*{\KNU}{Kyungpook National University, Daegu 702-701, Republic of Korea}
\newcommand*{\KNUindex}{29}
\newcommand*{\LPSC}{LPSC, Universite Joseph Fourier, CNRS/IN2P3, INPG, Grenoble, France}
\newcommand*{\LPSCindex}{30}
\newcommand*{\NSU}{Norfolk State University, Norfolk, Virginia 23504}
\newcommand*{\NSUindex}{31}
\newcommand*{\OHIOU}{Ohio University, Athens, Ohio  45701}
\newcommand*{\OHIOUindex}{32}
\newcommand*{\ODU}{Old Dominion University, Norfolk, Virginia 23529}
\newcommand*{\ODUindex}{33}
\newcommand*{\RPI}{Rensselaer Polytechnic Institute, Troy, New York 12180-3590}
\newcommand*{\RPIindex}{34}
\newcommand*{\URICH}{University of Richmond, Richmond, Virginia 23173}
\newcommand*{\URICHindex}{35}
\newcommand*{\ROMAII}{Universita' di Roma Tor Vergata, 00133 Rome Italy}
\newcommand*{\ROMAIIindex}{36}
\newcommand*{\MSU}{Skobeltsyn Nuclear Physics Institute, Skobeltsyn Nuclear Physics Institute, 119899 Moscow, Russia}
\newcommand*{\MSUindex}{37}
\newcommand*{\SCAROLINA}{University of South Carolina, Columbia, South Carolina 29208}
\newcommand*{\SCAROLINAindex}{38}
\newcommand*{\UNIONC}{Union College, Schenectady, NY 12308}
\newcommand*{\UNIONCindex}{39}
\newcommand*{\GLASGOW}{University of Glasgow, Glasgow G12 8QQ, United Kingdom}
\newcommand*{\GLASGOWindex}{40}
\newcommand*{\VIRGINIA}{University of Virginia, Charlottesville, Virginia 22901}
\newcommand*{\VIRGINIAindex}{41}
\newcommand*{\WM}{College of William and Mary, Williamsburg, Virginia 23187-8795}
\newcommand*{\WMindex}{42}
\newcommand*{\NOWMSU}{Skobeltsyn Nuclear Physics Institute, Skobeltsyn Nuclear Physics Institute, 119899 Moscow, Russia}
\newcommand*{\NOWINFNGE}{INFN, Sezione di Genova, 16146 Genova, Italy}
\newcommand*{\Rutg}{Physics \& Astronomy Department, Rutgers University, Piscataway, NJ 08855}
\newcommand*{\Syrac}{Physics Department, Syracuse University, Syracuse, NY 13244}
\newcommand*{\Virg}{Physics Department, University of Virginia, Charlottesville, VA 22904}

\author[toANL]{L. ~El Fassi\fnref{toRutg}} 
\author[toUNH,toSyrac]{L. ~Zana} 
\author[toANL]{K. ~Hafidi}
\author[toUNH]{M.~Holtrop} 
\author[toANL]{B. Mustapha} 
\author[toUTFSM,toJLAB]{W.K.~Brooks}
\author[toUTFSM,toYEREVAN]{H.~Hakobyan}
\author[toANL,toVirg]{X.~Zheng}
\author[toODU]{K.P. ~Adhikari}
\author[toODU]{D.~Adikaram}
% ...
 
 \address{\normalsize{(The CLAS Collaboration)}}
 \address[toANL]{\ANL} 
 \address[toUNH]{\UNH} 
 \address[toUTFSM]{\UTFSM} 
 \address[toJLAB]{\JLAB} 
 \address[toYEREVAN]{\YEREVAN} 
% ...
\address[toODU]{\ODU} 
% ...
 
 \fntext[toRutg]{Current address: Physics \& Astronomy Department, Rutgers University, Piscataway, NJ 08855}
 \fntext[toSyrac]{Current address: Physics Department, Syracuse University, Syracuse, NY 13244}
 \fntext[toVirg]{Current address: Physics Department, University of Virginia, Charlottesville, VA 22904}
 \fntext[toNOWMSU]{Current address: Skobeltsyn Nuclear Physics Institute, 119899 Moscow, Russia}
 \fntext[toNOWINFNGE]{Current address: 16146 Genova, Italy}

\date{\today}

\begin{abstract}
We have measured the nuclear transparency of the incoherent diffractive $A(e,e'\rho^0)$ process in $^{12}$C and $^{56}$Fe targets relative to $^2$H using a 5 GeV electron beam. The nuclear transparency, the ratio of the produced $\rho^0$'s on a nucleus relative to deuterium, which is sensitive to $\rho A$ interaction, was studied as function of the coherence length ($l_c$), a lifetime of the hadronic fluctuation of the virtual photon, and the four-momentum transfer squared ($Q^2$). While the transparency for both $^{12}$C and $^{56}$Fe showed no $l_c$ dependence, a significant $Q^2$ dependence was measured, which is consistent with calculations that included the color transparency effects.

\end{abstract}




\end{frontmatter}


\begin{linenumbers}

\begin{figure}[htp] 
\includegraphics[clip=true,trim= 0.825cm 8.5cm 0.9cm 9.25cm,width=0.525\textwidth]{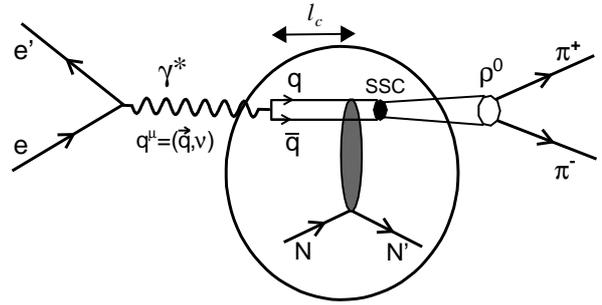}
\vglue -0.3cm 
\caption{An illustration of the creation of a SSC and its evolution to a fully formed $\rho^{0}$ (see the text for a full description).} 
\label{fig:reaction}
\vspace*{-0.25cm}
\end{figure}

\begin{figure}[t]
   \centering
  \includegraphics[clip=true,trim=1.275cm 0.8cm 0.5cm 2.55cm,width=0.485\textwidth]{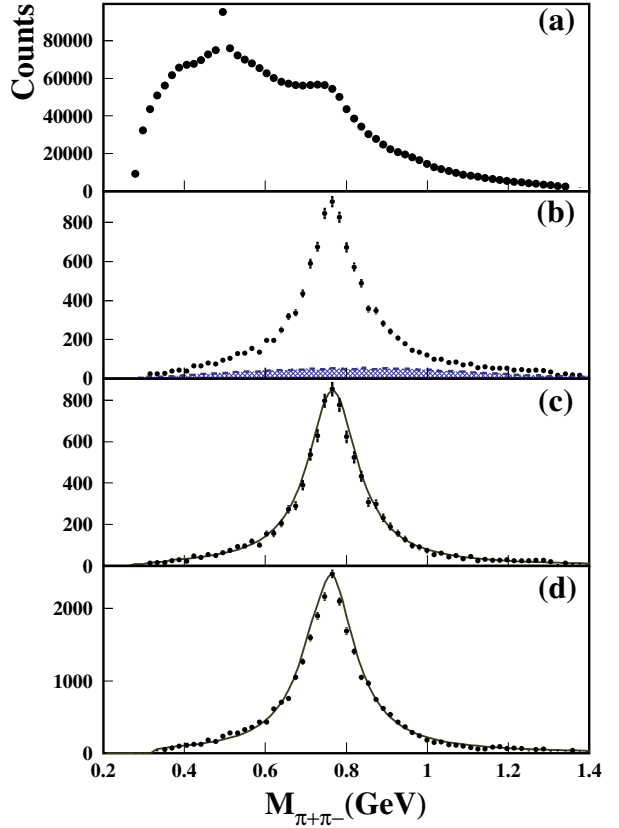}
   \vglue -0.85cm \caption{\textbf{(color online)} The (${\pi^+, \pi^-}$) invariant mass histogram for iron. Panel (a): Before applying kinematic cuts. Panel (b): after applying kinematic cuts. The blue shadow area represents the background contribution. Panel (c): after background subtraction. Panel (d): The (${\pi^+, \pi^-}$) invariant mass histogram for deuterium after background subtraction. The solid curves are non-relativistic Breit-Wigner fit to the data.}
   \label{fig:mass_spect} \vspace*{-0.25cm}
\end{figure}

Quantum chromodynamics (QCD) predicts that hadrons produced in exclusive reactions with sufficiently high squared four momentum transfers ($Q^2$) can pass through nuclear matter with dramatically reduced interactions \cite{Bertsch81,Zamolodchikov81,Brodsky88}. This is the so-called color transparency (CT) phenomenon, a key property of QCD as a color gauge theory. According to QCD, hard exclusive processes have the power to select special configurations of the hadron wave function where all quarks are close together, forming a color neutral small size configuration (SSC) with transverse size $r_\perp \sim 1/Q$. In these SSCs, the external color field vanishes as the distance between quarks shrinks and their color fields cancel each other, similar to the reduced electric field of a very small electric dipole. The reduced color field of the SSC allows it to propagate through a nucleus with little attenuation \cite{Jain96,Frankfurt94}.\\
\hspace*{0.25cm} 
Nuclear transparency, defined as the ratio of nuclear cross section per nucleon to that on a free nucleon, is the observable used to search for CT. The experimental signature of CT is the increase of the nuclear transparency, as $Q^2$ increases due to decreased interaction of the $\rho^0$ on its way out. In the absence of CT effects, the hadron-nucleon total cross-section, and thus the nuclear transparency, are nearly energy-independent \cite{Glauber59}.\\
\hspace*{0.25cm} 
Observation of CT requires that the SSC propagates a reasonable distance through the nucleus before expanding into a fully formed hadron. At very high energies, this is easily achieved since the SSC is highly relativistic and its lifetime in the nucleus rest frame will be dilated \cite{Brodsky88,Farrar88}. CT at high energy was observed by the E791 experiment \cite{Aitala01} at Fermilab which studied the A dependence of coherent diffractive dissociation of 500 GeV pions into di-jets on carbon and platinum targets. As predicted \cite{Frankfurt93,Frankfurt00}, the A-dependence of the measured cross sections was consistent with the nuclear target acting as a filter that removes all but small size configurations of the pion wave function. Due to the small transverse separation of the filtered $q\bar{q}$ pair, the quark and anti-quark each form a jet of hadrons in the final state. At low and intermediate energies, where the SSC travels a very short distance before evolving into a hadron, the situation is more challenging. In this kinematical region, the interaction of the hadron with the nucleus depends on the momentum at which it is produced, the evolution time of the SSC, its interaction cross section as it evolves into a normal state, and the distance it must travel through the nucleus.\\
\hspace*{0.25cm} 
Studying CT at low energies provides valuable information on SSC formation, expansion and, most importantly, its interaction as a function of its color field. CT is a key property of QCD. It offers a unique probe of ``color", a defining feature of QCD, yet totally invisible in the observed structure of ordinary nuclear matter. Establishing the kinematic conditions for the onset of CT is also critical to the future program of proton structure studies based on deep exclusive meson processes where the CT property of QCD is routinely used in the proof of QCD factorization theorem \cite{Strikman00}. Recently, CT was proposed \cite{Brodsky08} as the possible cause of the anomalous increase with centrality in the ratio of protons-to-pions produced at large transverse momenta in gold-gold collisions at the relativistic heavy ion collider in Brookhaven National Lab \cite{Adler03}.\\
\hspace*{0.25cm} 
Searches for CT with proton knock-out have all been negative \cite{Garino92,Makins94,O'Neill94,Abbott98,Garrow02} or inconclusive \cite{Carroll88,Mardor98,Leksanov01}, while results for meson production~\cite{Dutta03,Clasie07,Adams95,Airapetian03} have been more promising. The reason could be that the creation of a SSC is more probable for a meson  than for a baryon since only two quarks have to be localized to form the SSC. The first hint of CT at moderate energies was obtained in pion photoproduction off $^{4}$He \cite{Dutta03} with photon energies up to 4.5 GeV, but the experiment needed greater statistical precision to achieve conclusive findings. Another experiment~\cite{Clasie07} studied pion electroproduction off $^{12}$C, $^{27}$Al, $^{64}$Cu and $^{197}$Au over a range of $Q^{2}$ = 1.1 - 4.7 GeV$^{2}$. The nuclear transparencies of all targets relative to deuterium showed an increase with increasing $Q^{2}$. The most statistically significant result corresponds to the nuclear transparency for $^{197}$Au, which when fitted with a linear $Q^{2}$ dependence resulted in a slope of 0.012 $\pm$ 0.004 GeV$^{-2}$. The authors concluded that measurements at still higher momentum transfer would be needed to firmly establish the onset of CT.\\
\hspace*{0.25cm} 
Exclusive diffractive electroproduction of $\rho^{0}$ mesons provides a tool of choice to study color transparency. The advantage of using $\rho^{0}$ mesons is that they have the same quantum numbers as photons and so can be produced by a simple diffractive interaction, which selects small size initial state~\cite{Kopeliovich02}. In this process, illustrated in Fig.~\ref{fig:reaction}, the incident electron exchanges a virtual photon with the nucleus. The photon can then fluctuate into a virtual $q\bar{q}$ pair
 \cite{Bauer78} of small transverse separation~\cite{Brodsky94} proportional to $1/Q$, which can propagate over a distance $l_c=2\nu/(Q^{2}+M_{q\bar{q}}^2)$, known as the coherence length, where $\nu$ is the energy of the virtual photon and $M_{q\bar{q}}$ is the invariant mass of the $q\bar{q}$ pair. The virtual  $q\bar{q}$ pair can then scatter diffractively off a bound nucleon and becomes an on mass shell SSC. While expanding in size, the SSC travels through the nucleus and ultimately evolves to a fully formed $\rho^{0}$, which, in the final state, decays into a ($\pi^{+},\,\pi^{-}$) pair. By increasing $Q^{2}$, the size of the selected SSC can be reduced and consequently the nuclear transparency for the $\rho^{0}$ should increase.\\
\hspace*{0.25cm}
The nuclear transparency, $T_A$, is taken to be the ratio of the observed $\rho^{0}$ mesons per nucleon produced on a nucleus (A) relative to those produced from deuterium, where no significant absorption is expected. CT should yield an increase of $T_A$ with $Q^2$, but measurements by the HERMES~\cite{Ackerstaff99} collaboration show that $T_A$ also varies with $l_c$, which can also lead to a $Q^2$ dependence. Thus, to unambiguously identify CT, $l_c$ should be held constant or, alternatively, kept small compared to the nuclear radius to minimize the interactions of the $q\bar{q}$ pair prior to the diffractive production of the SSC.\\
\hspace*{0.25cm}
Fermilab experiment E665~\cite{Adams95} and the HERMES experiment~\cite{Airapetian03} at DESY used exclusive diffractive $\rho^{0}$ leptoproduction to search for CT. However, both measurements lacked the necessary statistical precision. HERMES measured the $Q^2$ dependence of the nuclear transparency for several fixed $l_c$ values. A simultaneous fit of the $Q^2$ dependence over all $l_c$ bins resulted in a slope of 0.089 $\pm$ 0.046 GeV$^{-2}$. The unique combination of high beam intensities available at the Thomas Jefferson National Accelerator Facility know as JLab and the wide kinematical coverage provided by the Hall B large acceptance spectrometer \cite{Mecking03} (CLAS) was key to the success of the measurements reported here.\\
\hspace*{0.25cm} 
The experiment ran during the winter of 2004. An electron beam with 5.014 GeV energy was incident simultaneously on a 2 cm liquid deuterium target and a 3 mm diameter solid target (C or Fe). The nuclear targets were chosen to optimize two competing requirements; provide sufficient nuclear path length compared to the SSC expansion length while minimizing the probability of $\rho^0$ decay inside the nucleus. A new double-target system~\cite{Hakobyan08} was developed to reduce systematic uncertainties and allow high precision measurements of the transparency ratios between heavy targets and deuterium. The cryogenic and solid targets were located 4 cm apart to minimize the difference in CLAS acceptance while maintaining the ability to identify the target where the interaction took place event-by-event via vertex reconstruction. The thickness of the solid targets (1.72 mm for carbon and 0.4 mm for iron) were chosen so that all of the targets including deuterium had comparable luminosities ($\sim$ 10$^{34}$ nucleon cm$^{-2}$ s$^{-1}$). The scattered electrons and two oppositely charged pions were detected in coincidence using the CLAS spectrometer. The scattered electrons were identified using the \v Cerenkov and the electromagnetic calorimeter while the pions were identified through time-of flight measurements \cite{Mecking03}.\\
\hspace*{0.25cm}
The $\rho^0$ mesons were identified through the reconstructed invariant mass of the two detected pions with 0.6 $< M_{\pi^+ \pi^-} < $ 1 GeV. For each event, several kinematic variables were evaluated including $Q^2$, $l_c$ using the $\rho^0$ mass instead of $M_{q\bar{q}}$, the photon-nucleon invariant mass squared $W^2$, the squared four-momentum transfer to the target $t$, and the fraction of the virtual photon energy carried by the $\rho^0$ meson $z_\rho = E_\rho/\nu$ where $E_\rho$ is the energy of the $\rho^0$. To identify exclusive diffractive and incoherent $\rho^0$ events, a set of kinematic conditions had to be satisfied. We required $W > $ 2 GeV to suppress pions from decay of resonances, $-t <$ 0.4 GeV$^2$ to select diffractive events, $-t >$ 0.1 GeV$^2$ to exclude coherent production off the nucleus and $z_\rho >$ 0.9 \textbf{t}o select elastically produced $\rho^0$ mesons. The two pions invariant mass distributions are shown in Fig.~\ref{fig:mass_spect}. After applying all the cuts, the invariant mass distribution (Fig\ref{fig:mass_spect}.b) exhibits a clean $\rho^0$ peak positioned around 770 MeV with the expected width of 150 MeV. A good description of the data was obtained using our Monte-Carlo (MC) simulation. Our generator simulates the $\rho^0$ electroproduction process and the main channels that may produce a ($\pi^+, \pi^-$) pair in the final state and contribute to the background underneath the $\rho^0$ peak. These channels are $ep \rightarrow e \Delta^{++} \pi^-$, $ep \rightarrow e \Delta^{0} \pi^+$ and a non resonant $ep \rightarrow e p \pi^+ \pi^-$. The cross sections of these processes were taken from existing measurements~\cite{Cassel81}. The standard CLAS GEANT based simulation packages was used to simulate the experimental apparatus. The Fermi motion of the nucleons in nuclei was simulated by folding the elementary cross section with the spectral function of the target using a realistic model~\cite{Ciofi96}. Radiative effects are also included in the simulation.\\ 
\hspace*{0.25cm}
The nuclear transparency for a given target, with nucleon number A, is defined as
\end{linenumbers} 
\begin{equation}
T_A = (N^\rho_A/\mathcal{L}^{int}_A) /(N^\rho_D/\mathcal{L}^{int}_D),
\end{equation}
\begin{linenumbers}
where $D$ refers to deuterium, and $A$ to carbon or iron, $\mathcal{L}^{int}_{A, D}$  to the integrated luminosities and $N^\rho_{A, D}$ to the number of incoherent $\rho^0$ events per nucleon after subtraction of background contributions.~The transparency ratios were also corrected from detector and reconstruction efficiencies,~acceptance and radiative effects, Fermi motion and contributions from the liquid deuterium target windows.~The CLAS acceptance and reconstruction efficiencies were evaluated with the simulations described earlier.~Data from both simulation and measurements were processed with the same analysis code.~Based on the comparison between data and MC, the acceptance was defined in each elementary bin in all relevant variables; Q$^2$, $t$, $W$, the $\rho^0$ momentum $P_{\rho^0}$, and the decay angle in the $\rho^0$ rest frame $\theta_{\pi^+}$, as the ratio of accepted to generated events. Each event was then weighted with the inverse of the corresponding acceptance.~The weighted ($\pi^+, \pi^-$) mass spectra were fitted as shown in Fig.~\ref{fig:mass_spect}.c using a non relativistic Breit-Wigner for the shape of a $\rho^0$ while the shape of
\begin{figure}[htp]
\vglue -0.7cm
   \centering
  \includegraphics[clip=true,trim=0.9cm 0.7cm -3.7cm 1.475cm,width=0.77\textwidth]{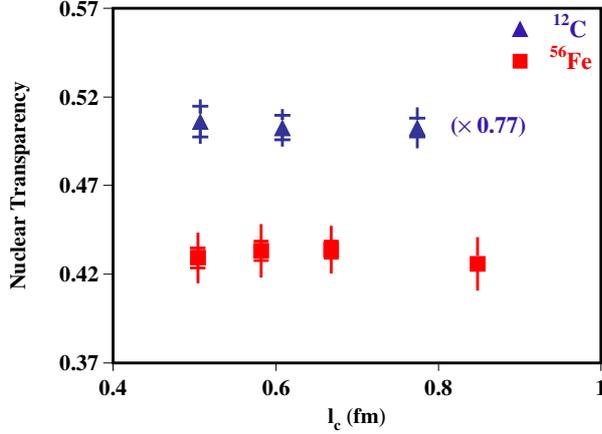}
\vglue -1.465cm \caption{\textbf{(color online)} Nuclear transparency as a function of $l_c$. The inner error bars are the statistical uncertainties and the outer ones are the statistical and point-to-point ($l_c$ dependent) systematic uncertainties added in quadrature. There is an additional normalization systematic uncertainty of 1.9\% for carbon and 1.8\% for iron (not shown in the figure) with acceptance and background subtraction being the main sources. The carbon data has been scaled by a factor 0.77 to fit in the same figure with the iron data.}  
\label{fig:lc_results} \vspace*{-0.3cm}
\end{figure} the background was taken from the simulation.~The magnitudes of each contributing process were taken as free parameters in the fit of the mass spectra.~The acceptance correction to the transparency ratio was found to vary between 5 and 30\%.~Radiative corrections were extracted for each ($l_c$, Q$^2$) bin using our MC generator in conjunction with the DIFFRAD~\cite{Akushevich99} code developed for exclusive vector meson production.~The radiative correction to the transparency ratio was found to vary between 0.4 and 4\%. An additional correction of around 2.5\% was applied to account for the contribution of deuterium target endcaps.~The corrected $t$ distributions for exclusive events were fit with an exponential form $Ae^{-bt}$.~The slope parameters $b$ for $^2$H (3.59 $\pm$ 0.5), C (3.67 $\pm$ 0.8) and Fe (3.72 $\pm$ 0.6) were reasonably consistent with CLAS~\cite{Morrow08} hydrogen measurements of 2.63 $\pm$ 0.44 taken with 5.75 GeV beam energy.\\
\hspace*{0.25cm} 
The transparencies for C and Fe are shown as a function of $l_c$ in Fig.~\ref{fig:lc_results}. As expected, they do not exhibit any $l_c$ dependence because $l_c$ is much shorter than the C and Fe nuclear radii of 2.7 and 4.6 fm respectively. 
\begin{figure}[t]
  \includegraphics[clip=true,trim=0.7cm 0.5cm -2.55cm 2.15cm,width=0.65\textwidth]{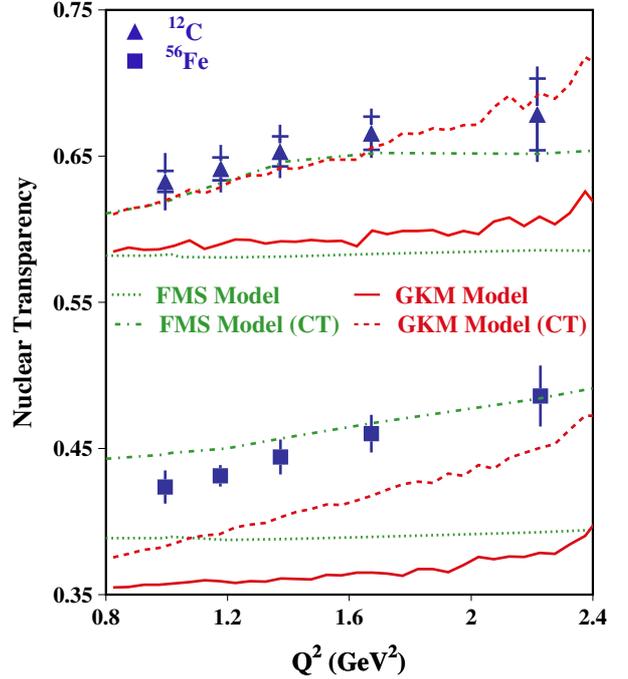}
  \vglue -0.905cm \caption{\textbf{(color online)} Nuclear transparency as a function of $Q^2$. The inner error bars are statistic uncertainties and the outer ones are statistic and point-to-point ($Q^2$ dependent) systematic uncertainties added in quadrature. The curves are predictions of the FMS \cite{Frankfurt08} (red) and GKM \cite{Gallmeister10} (green) models with (dashed-dotted and dashed curves, respectively) and without (dotted and solid curves, respectively) CT. Both models include the pion absorption effect when the $\rho^0$ meson decays inside the nucleus. There is an additional normalization systematic uncertainty of 2.4\% for carbon and 2.1\% for iron (not shown in the figure).}
  \label{fig:ct_results} \vspace*{-0.2cm}
\end{figure} Consequently, the coherence length effect cannot mimic the CT signal in this experiment.\\
\hspace*{0.25cm} 
Fig.~\ref{fig:ct_results} shows the increase of the transparency with $Q^2$ for both C and Fe. The data are consistent with expectations of CT.~Note that in the absence of CT effects, hadronic Glauber calculations would predict no $Q^2$ dependence of $T_A$ since any $Q^2$ dependence in the $\rho^0$ production cross section would cancel in the ratio.~The rise in transparency with $Q^2$ corresponds to an $(11 \pm 2.3)\%$ and $(12.5 \pm 4.1)\%$ decrease in the absorption of the $\rho^0$ in Fe and C respectively. The systematics uncertainties were separated into point-to-point uncertainties, which are $l_c$ dependent in Fig.~\ref{fig:lc_results} and Q$^2$ dependent in Fig.~\ref{fig:ct_results} and normalization uncertainties, which are independent of the kinematics. Effects such as kinematic cuts, model dependence in the acceptance correction and background subtraction, Fermi motion and radiative correction were studied and taken into account in the systematic uncertainties described in details in \cite{elfassi12}. The fact that we were able to observe the increase in nuclear transparency requires that the SSC propagated sufficiently far in the nuclear medium and experienced reduced interaction with the nucleons before evolving to a normal hadron. The $Q^2$ dependence of the transparency was fitted by a linear form $T_A = a~Q^2 + b$. The extracted slopes ``$a$'' for C and Fe are compared to the model predictions in Table~\ref{tab:slopes}. Our results for Fe are in good agreement with both Kopeliovich-Nemchik-Schmidt (KNS)~\cite{Kopeliovich07} and Gallmeister-Kaskulov-Mosel (GKM)~\cite{Gallmeister10} predictions, but somewhat larger than the Frankfurt-Miller-Strikman (FMS)~\cite{Frankfurt08} calculations.~While the KNS and GKM models yield an approximately linear $Q^2$ dependence, the FMS calculation yields a more complicated $Q^2$ dependence as shown in Fig.~\ref{fig:ct_results}. The measured slope for carbon corresponds to a drop in the absorption of the $\rho^0$ from 37\% at $Q^2$ = 1 GeV$^2$ to 32\% at $Q^2$ = 2.2 GeV$^2$, in reasonable agreement with the calculations. Despite the differences between these models in the assumed production mechanisms and SSC interaction in the nuclear medium, they all support the idea that the observed $Q^2$ dependence is clear evidence for the onset of CT, demonstrating the creation of small size configurations, their relatively slow expansion and their reduced interaction with the nuclear medium.\\
\begin{table}[t]
 \vglue -0.15cm\caption{\label{tab:table1}Fitted slope parameters of the $Q^2$-dependence of the nuclear transparency for carbon and iron nuclei.~The results are compared with theoretical predictions of KNS~\cite{Kopeliovich07}, GKM~\cite{Gallmeister10} and FMS~\cite{Frankfurt08}.} \vspace*{0.25cm}
\begin{tabularx}{225pt}{XXXXX}
\hline\hline
&\multicolumn{1}{X}\small{Measured Slopes}&\multicolumn{3}{X}\small{Model Predictions}\\
\small{Nucleus}&\hspace*{1.5cm}GeV$^{-2}$&\hspace*{0.6cm}\small{KNS}&\hspace*{0.6cm}\small{GKM}&\hspace*{0.7cm}\small{FMS}\\ \hline
\centering
\hspace*{0.15cm}\small{C}&\hspace*{0.25cm}\small{$\mathrm{0.044\pm0.015_{stat}\pm0.019_{syst}}$}&\hspace*{0.65cm}\small{0.06}&\hspace*{0.65cm} \small{0.06} &\hspace*{0.65cm} \small{0.025} \\
\centering \hspace*{0.15cm}\small{Fe}&\hspace*{0.25cm}\small{$\mathrm{0.053\pm0.008_{stat}\pm0.013_{syst}}$}&\hspace*{0.6cm}\small{0.047}&\hspace*{0.6cm} \small{0.047}&\hspace*{0.7cm}\small{0.032} \\
 \hline\hline 
\end{tabularx}
\label{tab:slopes}
\end{table} 
\hspace*{0.25cm}
The onset of CT in $\rho^0$ electroproduction seems to occur at lower $Q^2$ than in the pion measurements. This early onset suggests that diffractive meson production is the optimal way to create a SSC \cite{Kopeliovich02}. The $Q^2$ dependence of the transparency ratio is mainly sensitive to the reduced interaction of the SSC as it evolves into a full-sized hadron, and thus depends strongly on the expansion length over which the SSC color fields expand to form a $\rho^0$ meson. The expansion length used by the FMS and GKM models is between 1.1 and 2.4 fm for $\rho^0$ mesons produced with momenta from 2 to 4.3 GeV while the KNS model uses an expansion length roughly a factor of two smaller. The agreement between the observed $Q^2$ dependence and these models suggests that these assumed expansion distances are reasonable, yielding rest-frame SSC lifetimes of about $0.5 - 1 \times10^{-24}$ second.\\
\hspace*{0.25cm}
In summary, we have experimentally observed the formation of small size configurations in diffractive $\rho^0$ meson electroproduction and its reduced interaction as it travels through the nucleus. Our data are consistent with expectations of color transparency and, based on the existing models, provide the first estimate of the expansion time (lifetime) for these exotic configurations. Having established these features, detailed studies of the theoretical models will allow the first quantitative evaluation of the structure and evolution properties of the SSCs. Such studies will be further enhanced by future measurements \cite{Hafidi06}, which will include additional nuclei and extend to higher $Q^2$ values.\\

We thank M.~Sargsian for independent evaluation of the Fermi
motion effect. We would like to acknowledge the outstanding efforts of the staff of the Accelerator and the Physics Divisions at Jefferson Lab that made this experiment possible. This work was
supported in part by the U.S. Department of Energy, Office of Nuclear Physics, under
contracts No.~DE-AC02-06CH11357 and No.~DE-FG02-88ER40410, the Chilean FONDECYT grants 1080564 and 3100064 CONICYT ACT-119, BASAL FB0821, the National Science Foundation, the Italian Istituto Nazionale di Fisica Nucleare, the French Centre National de la Recherche Scientifique, the French Commissariat \`{a}
l'Energie Atomique, the National Research Foundation of Korea, the UK Science and
Technology Facilities Council (STFC), and the Physics Department at Moscow State
University. The Jefferson Science Associates (JSA) operates the Thomas Jefferson
National Accelerator Facility for the United States Department of Energy under contract DE-AC05-06OR23177.

\bibliographystyle{model1-num-names}

\end{linenumbers}

\end{document}